\documentclass[12pt,a4paper]{article}
\usepackage[utf8]{inputenc}
\usepackage{epsf,amsmath,amssymb,graphicx,dcolumn}
\usepackage{caption}
\usepackage{subcaption}
\usepackage{scalefnt,ulem,pstricks}
\usepackage{booktabs,multirow,tabularx}
\usepackage{cite,hyperref}
\usepackage{cleveref}
\usepackage{color}
\usepackage{rotating}
\usepackage{microtype}
\usepackage[titletoc,title]{appendix}
\usepackage[numbers,sort&compress]{natbib}
\usepackage{amsmath,amsfonts,amsthm,bm} % Math packages
\usepackage[tikz]{bclogo}
\usepackage{subcaption}
\usepackage{tikz-feynman}
 
\newcommand{\sphid}[1]{}

\providecommand{\href}[2]{#2}

\newcommand\as{\alpha_{\mathrm{S}}}

\def\to{\rightarrow}

\def\ptveto{p_{T,\ell\ell\gamma}^{\rm veto}}

\newcommand\Matrix{{\sc Matrix}}
\newcommand\Munich{{\sc Munich}}
\newcommand\OpenLoops{{\sc OpenLoops 2}}

\newcommand{\qt}{\ensuremath{q_T}}
\newcommand{\pt}{\ensuremath{p_T}}

\newcommand{\eqn}[1]{Eq.\,(\ref{#1})}

\newcommand{\fig}[1]{Figure~\ref{#1}}

\newcommand{\tab}[1]{Table~\ref{#1}}

\def\citere#1{\mbox{Ref.~\cite{#1}}}
\def\citeres#1{\mbox{Refs.~\cite{#1}}}

\newcommand{\ptg}{\ensuremath{p_{T,\gamma}}}
\newcommand{\ptllg}{\ensuremath{p_{T,\ell\ell\gamma}}}

\newcommand{\ptlone}{\ensuremath{p_{T,\ell_1}}}
\newcommand{\ptltwo}{\ensuremath{p_{T,\ell_2}}}

\newcommand{\etal}{\ensuremath{\eta_{\ell}}}

\newcommand{\rd}{\mathrm{d}} % differential operator

\usepackage{etoolbox}
\makeatletter
\patchcmd{\@sect}{#8}{\boldmath #8}{}{}
\let\ori@chapter\@chapter
\def\@chapter[#1]#2{\ori@chapter[\boldmath#1]{\boldmath#2}}
\makeatother

\begin{document} 
\begin{flushright}
\vspace*{-1.5cm}
MPP-2020-99\
\end{flushright}
\vspace{0.cm}

\begin{center}
{\Large \bf The $\bm{Z\gamma}$ transverse-momentum spectrum at NNLO+N{\boldmath $^3$}LL}
\end{center}

\begin{center}
{\bf Marius Wiesemann$^{(a)}$}, {\bf Luca Rottoli}$^{(b),(c)}$, and {\bf Paolo Torrielli$^{(d)}$}

$^{(a)}$ Max-Planck-Institut f\"ur Physik, F\"ohringer Ring 6, 80805 M\"unchen, Germany\\
$^{(b)}$ Dipartimento di Fisica G. Occhialini, Universit\`a degli Studi di Milano-Bicocca and INFN, Piazza della Scienza 3, 20126 Milano, Italy\\
$^{(c)}$ Ernest Orlando Lawrence Berkeley National Laboratory, University of California, Berkeley, CA 94720, USA\\
$^{(d)}$ Dipartimento di Fisica and Arnold-Regge Center, Universit\`a di Torino and INFN, Sezione di
Torino, Via P. Giuria 1, I-10125, Turin, Italy

\href{mailto:marius.wiesemann@cern.ch}{\tt marius.wiesemann@cern.ch}\\
\href{mailto:luca.rottoli@unimib.it}{\tt luca.rottoli@unimib.it}\\
\href{mailto:torriell@to.infn.it}{\tt torriell@to.infn.it}

\end{center}

\begin{center} {\bf Abstract} \end{center}\vspace{-1cm}
\begin{quote}
\pretolerance 10000

We consider the transverse-momentum (\pt{}) distribution of $Z\gamma$ pairs produced in hadronic collisions. 
Logarithmically enhanced contributions at small \pt{} are resummed to all orders in QCD perturbation theory and combined with the fixed-order prediction.
We achieve the most advanced prediction for the $Z\gamma$ \pt{} spectrum 
by matching next-to-next-to-next-to-leading logarithmic (N$^3$LL) resummation
to the integrated cross section at next-to-next-to-leading order (NNLO). 
By considering $\ell^+\ell^-\gamma$ production at the fully differential level, including spin correlations, interferences and off-shell effects, arbitrary cuts can be applied to the leptons and the photon.
We present results at the LHC
in presence of fiducial cuts and find agreement with the $13$\,TeV ATLAS data at the few-percent level.
\end{quote}

\parskip = 1.2ex 

Vector-boson pair production processes are an integral part of the rich physics programme at the Large Hadron Collider (LHC). 
They play a crucial role in both precision measurements of Standard-Model (SM) rates and the search for new-physics phenomena. 
In particular, production processes of neutral vector bosons, like $Z\gamma$ production, provide very clean experimental signatures in
the $Z\to \ell^+\ell^-$ decay channels, since the final state can be fully reconstructed. 
Their pure experimental signatures and relatively large cross sections render them well suited to search for anomalous couplings. 
For instance, the measurement of a non-zero $ZZ\gamma$ coupling, which is absent in the SM, 
would be direct evidence of physics beyond the SM (BSM). $Z\gamma$ production contributes also as irreducible background to direct searches 
for BSM resonances and to Higgs boson measurements, see e.g.\ \citere{Aad:2020plj}. Although the decay into a $Z\gamma$ pair of the Higgs boson is a rare 
loop-induced process in the SM, new-physics extensions may significantly enhance this decay channel. 
%See \citere{Aad:2020plj} for instance for a recent search in this channel.

The precise knowledge of rates and distributions in $Z\gamma$ production provides 
a strong test of the gauge structure of electroweak (EW) interactions and the mechanism of EW symmetry breaking. 
Measurements of $Z\gamma$ production have been carried out at the LHC 
at 7\,TeV~\cite{Chatrchyan:2011rr,Aad:2011tc,Aad:2012mr,Chatrchyan:2013fya,Chatrchyan:2013nda,Aad:2013izg},
8\,TeV~\cite{Aad:2014fha,Khachatryan:2015kea,Khachatryan:2016yro,Aad:2016sau}, 
and 13\,TeV~\cite{Aaboud:2018jst,Aad:2019gpq}. 
The latest measurement of \citere{Aad:2019gpq} is the first diboson analysis to use the full Run~II data set and achieves remarkably small experimental uncertainties.
%Also searches for new heavy $Z\gamma$ resonances involving both charged leptons and neutrinos have been performed, see \citere{} for example.

To match the precision achieved by the experiments a significant effort has been made to advance theoretical predictions 
for $Z\gamma$ production in the past years.
The next-to-leading order (NLO) QCD cross section has been known for some time both for on-shell $Z$ bosons \cite{Ohnemus:1992jn} and including their leptonic decays \cite{Baur:1997kz}.
The loop induced gluon-fusion contribution was the first contribution to the next-to-next-to-leading order (NNLO) QCD cross section to be computed \cite{Ametller:1985di,vanderBij:1988fb,Adamson:2002rm}.
In \citere{Campbell:2011bn} the NLO cross section, including photon radiation off the leptons, and the loop-induced gluon fusion contribution were combined.
The complete NNLO QCD corrections to $\ell^+\ell^-\gamma$ production at the fully differential level were first calculated in \citeres{Grazzini:2013bna,Grazzini:2015nwa} and later confirmed by an 
independent calculation \cite{Campbell:2017aul}. Electroweak (EW) corrections were presented in \citeres{Hollik:2004tm,Accomando:2005ra}.

Two different mechanisms are relevant to produce isolated photons in 
the final state: a perturbative one through {\it direct} production in the underlying hard subprocess, and a non-perturbative one through {\it fragmentation} of a quark or a gluon.
The latter production mechanism requires the knowledge of the respective fragmentation functions to absorb singularities related to collinear photon emissions, and those functions are determined from data with relatively large uncertainties.
In experimental analyses the fragmentation component is typically suppressed 
by the criteria used to isolate photons.
On the theoretical side, the separation between the 
two production mechanisms is delicate, 
as sharply isolating the photon from the partons would spoil infrared (IR) safety.
Remarkably, by exploiting Frixione smooth-cone photon isolation~\cite{Frixione:1998jh} the fragmentation component can be completely removed in an IR-safe manner, which has the further advantage of substantially simplifying theoretical calculations of photon processes beyond the leading order (LO). 
Experimentally, the finite granularity of the 
calorimeter prevents a complete implementation of the smooth-cone isolation. As a consequence, experimental analyses 
rely instead on isolation criteria with a fixed cone. To facilitate data--theory 
comparisons, the smooth-cone parameters are typically tuned in comparisons with
calculations including fragmentation functions in order to mimic 
the fixed-cone isolation criteria of the experiments, see e.g. \citere{Catani:2018krb}.
Due to the large scale separation between the photon energy and the hadronic energy within the isolation cone, the presence of isolation cuts induces potentially large non-global (NG) logarithms, whose resummation is known up to leading-logarithmic (LL) accuracy \cite{Dasgupta:2001sh,Balsiger:2018ezi}.

In this paper we consider the transverse-momentum (\pt{}) distribution of $Z\gamma$ pairs. This distribution is
among the most important differential observables in $Z\gamma$ production, and it has recently been measured 
at a precision of a few percent by using the full Run~II data set \cite{Aad:2019gpq}. 
For the first time, we perform transverse-momentum resummation of $Z\gamma$ pairs at next-to-next-to-next-to-leading logarithmic 
(N$^3$LL) accuracy and match it to the NNLO integrated cross section. 
To this end, we calculate the process $pp\to \ell^+\ell^-\gamma$
with off-shell effects and spin correlations 
by consistently including all resonant and non-resonant topologies.
Our computation is fully differential in the momenta of the final-state leptons and the photon, which 
allows us to apply arbitrary fiducial cuts. 
%Phenomenological predictions of the $Z\gamma$ \pt{} spectrum at NNLO+N$^3$LL are compared the recent Run~II measurement at $13$\,TeV of \citere{}.

We employ the \textsc{Matrix+RadISH} interface \cite{Kallweit:2020gva}, which combines NNLO calculations within \textsc{Matrix} \cite{Grazzini:2017mhc,Matrixurl} with
the \textsc{RadISH} resummation formalism of \citeres{Monni:2016ktx,Bizon:2017rah,Monni:2019yyr}.
All tree-level and one-loop amplitudes are evaluated with \OpenLoops{} \cite{Cascioli:2011va,Buccioni:2017yxi,Buccioni:2019sur}.
At two-loop level we use the $q\bar{q}\to V\gamma$ amplitudes of \citere{Gehrmann:2011ab}. 
NNLO accuracy is achieved by a fully general implementation of the \qt{}-subtraction formalism \cite{Catani:2007vq}
within \Matrix{}. The NLO parts therein (for $Z\gamma$ and $Z\gamma$+$1$-jet)
are calculated by \Munich{}\footnote{The Monte
Carlo program \Munich{} features a general implementation of an
efficient, multi-channel based phase-space integration and computes
both NLO QCD and NLO EW~\cite{Kallweit:2014xda,Kallweit:2015dum} corrections
to arbitrary SM processes.}~\cite{munich}, which uses the 
Catani--Seymour dipole subtraction method \cite{Catani:1996jh,Catani:1996vz}.
The \Matrix{} framework features NNLO QCD corrections to a large number of colour-singlet processes at hadron colliders. It has already been used to obtain several 
state-of-the-art NNLO QCD predictions \cite{Grazzini:2013bna,Grazzini:2015nwa,Cascioli:2014yka,Grazzini:2015hta,Gehrmann:2014fva,Grazzini:2016ctr,Grazzini:2016swo,Grazzini:2017ckn,Kallweit:2018nyv}\footnote{It was also used in the NNLO+NNLL computation of \citere{Grazzini:2015wpa}, and in the NNLOPS computations of \citeres{Re:2018vac,Monni:2019whf,Alioli:2019qzz,Monni:2020nks}.}, and for massive diboson processes
it has been extended to combine NNLO QCD with NLO EW corrections \cite{Kallweit:2019zez} and with NLO QCD corrections to the loop-induced gluon fusion contribution \cite{Grazzini:2018owa,Grazzini:2020stb}. Through the recently implemented \textsc{Matrix+RadISH} interface \cite{Kallweit:2020gva} it is now also possible to deal with
the resummation of transverse observables such as the transverse momentum of the colour-singlet final state.
  
\begin{figure}[t]
  \begin{center}
\hspace{-0.4cm}
    \begin{subfigure}[b]{.33\linewidth}
      \centering
\begin{tikzpicture}
  \begin{feynman}
    \vertex (a1) {\( q\)};
    \vertex[below=1.6cm of a1] (a2){\(\overline q\)};
    \vertex[right=1.5cm of a1] (a3);
    \vertex[right=1.5cm of a2] (a4);
    \vertex[right=1.5cm of a3] (a5){\(\gamma\)};
    \vertex[right=2cm of a3] (a9);
    \vertex[right=1cm of a4] (a6);
    \vertex[below=0.7cm of a9] (a7){\(e^+\)} ;
    \vertex[below=1.7cm of a9] (a8){\(e^-\)};
    
    \diagram* {
      {[edges=fermion]
        (a1)--(a3)--(a4)--(a2),
        (a7)--(a6)--(a8),
      },
      (a3) -- [ boson] (a5),
      (a4) -- [boson, edge label'=\(Z/\gamma^*\)] (a6),
       };

  \end{feynman}
\end{tikzpicture}
\caption{}
        \label{subfig:t}
\end{subfigure}%
\begin{subfigure}[b]{.33\linewidth}
  \centering
\begin{tikzpicture}
  \begin{feynman}
    \vertex (a1) {\(  q\)};
    \vertex[below=1.6cm of a1] (a2){\(\overline q\)};
    \vertex[below=0.8cm of a1] (a3);
    \vertex[right=1.5cm of a3] (a4);
    \vertex[right=1cm of a4] (a5);
    \vertex[right=0.4cm of a5] (a6);
    \vertex[right=1.1cm of a6] (a7);
    \vertex[below=0.8cm of a7](a8){\(e^{+}\)};
    \vertex[above=0.5cm of a6](a9);
    \vertex[above=0cm of a7] (a10){\(\gamma\)}; ;
    \vertex[above=0.6cm of a7] (a11){\(e^{-}\)};
 
    \diagram* {
      {[edges=fermion]
        (a1)--(a4)--(a2),
        (a8)--(a5)--(a9)--(a11),
      },
      (a4) -- [boson, edge label=\(Z/\gamma^*\)] (a5),
      (a9) -- [boson] (a10),
       };

  \end{feynman}

\end{tikzpicture}
\caption{}
        \label{subfig:s}
\end{subfigure}
\begin{subfigure}[b]{.33\linewidth}
  \centering
\begin{tikzpicture}
  \begin{feynman}
    \vertex (a1) {\( g\)};
    \vertex[below=1.6cm of a1] (a2){\(g\)};
    \vertex[right=1.5cm of a1] (a3);
    \vertex[right=1.5cm of a2] (a4);
    \vertex[right=1cm of a3] (a5);
    \vertex[right=1cm of a4] (a6);
    \vertex[right=1.5cm of a5] (a7){\(\gamma\)};
    \vertex[right=1.9cm of a5] (a11);
    \vertex[right=0.9cm of a6] (a8);
    \vertex[below=0.7cm of a11] (a9){\(e^+\)} ;
    \vertex[below=1.7cm of a11] (a10){\(e^-\)};
 
    \diagram* {
      {[edges=fermion]
        (a3)--(a4)--(a6)--(a5)--(a3),
        (a9)--(a8)--(a10),
      },
      (a1) -- [ gluon] (a3),
      (a2) -- [ gluon] (a4),
      (a5) -- [ boson] (a7),
      (a6) -- [boson, edge label'=\(Z/\gamma^*\)] (a8),
       };

  \end{feynman}
\end{tikzpicture}
\caption{}
        \label{subfig:gg}
\end{subfigure}
\end{center}
\caption{\label{fig:diag} Feynman diagrams for the production of 
two charged leptons and a photon: (a-b) sample tree-level diagrams in the quark-annihilation channel contributing at LO; (c) sample loop-induced diagram in the gluon-fusion channel contributing at NNLO.}
\end{figure}
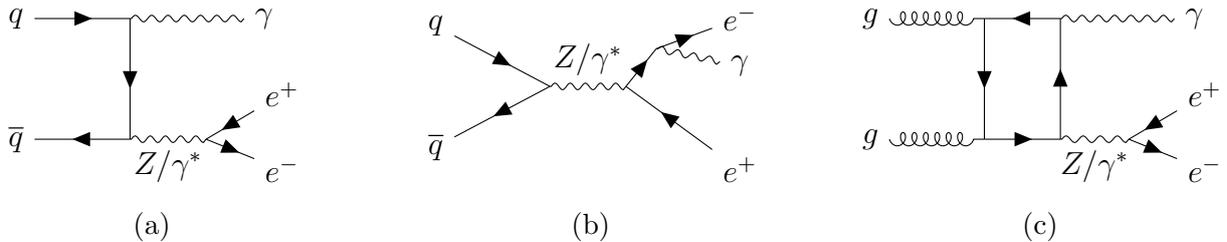

We consider the process 
\begin{align}
pp \rightarrow \ell^+\ell^-\,\gamma+X\nonumber
\end{align}
for massless leptons $\ell\in\{e,\mu\}$. Although our calculation also applies to the process
$pp \rightarrow \nu\bar\nu\,\gamma+X$, we do not consider it here, as the transverse momentum
of the $Z\gamma$ pair in that case cannot be experimentally reconstructed.
Representative LO diagrams are shown in \fig{fig:diag}\,(a-b). They are driven by quark annihilation in the initial state and involve single-resonant $t$-channel 
$Z\gamma$ production~(panel (a)) and single-resonant $s$-channel Drell--Yan (DY) topologies~(panel (b)).
\fig{fig:diag}\,(c) shows a loop-induced diagram that is driven by gluon fusion in the initial state and
enters the cross section at NNLO. 
The loop-induced gluon-fusion contribution is effectively only LO accurate and has 
Born kinematics. Therefore, it contributes trivially to the $Z\gamma$ transverse-momentum ($\ptllg$) distribution. Furthermore, its contribution  is rather small, being less than $10\%$ of the NNLO corrections and well below $1\%$ of 
the full $Z\gamma$ cross section at NNLO \cite{Grazzini:2017mhc}.
We thus refrain from including the loop-induced gluon-fusion contribution in our calculation.

The perturbative description of the $Z\gamma$ transverse-momentum 
spectrum at fixed order breaks down in kinematic regimes dominated by soft and collinear QCD radiation, i.e.\ at small \ptllg{}, due to the presence of large logarithms $L = \ln (m_{\ell\ell\gamma}/\ptllg)$, with $m_{\ell\ell\gamma}$ being the invariant mass of the $Z\gamma$ pair. 
Over the last four decades, a variety of formalisms has been developed to perform the resummation of large logarithmic contributions in the transverse momentum \pt{} of colour-singlet processes~\cite{Parisi:1979se,Collins:1984kg,Idilbi:2005er,Bozzi:2005wk,Catani:2010pd,Becher:2010tm,Becher:2011xn,GarciaEchevarria:2011rb,Chiu:2012ir,Monni:2016ktx,Ebert:2016gcn,Bizon:2017rah}.
We employ the \textsc{RadISH} formalism of \citeres{Monni:2016ktx,Bizon:2017rah,Monni:2019yyr} to resum the relevant logarithmic terms to  all orders. The logarithmic accuracy is customarily defined in 
terms of the logarithm of the cumulative cross section $\ln \sigma (\pt)$.
The dominant terms $\as^n L^{n+1}$ are referred to as leading logarithmic, terms of $\as^n L^{n}$ as 
next-to-leading logarithmic (NLL), terms of $\as^n L^{n-1}$ as next-to-next-to-leading logarithmic (NNLL), 
and so on. We perform the resummation of the $Z\gamma$ \pt{} spectrum up to N$^3$LL
based on the formul\ae\ presented in \citere{Bizon:2017rah}. 
The resummation formalism has been implemented in the \textsc{RadISH} code for Higgs and Drell-Yan production.
The application to more complex colour-singlet processes, such as $Z\gamma$ production, is 
achieved through the \textsc{Matrix+RadISH} interface \cite{Kallweit:2020gva}.
We note that the LL resummation of the loop-induced gluon-fusion contribution to the $Z\gamma$ cross section is formally of the same order as 
N$^3$LL corrections to the $q \bar q$ channel, and that both contributions can be treated completely independently. 
The proper treatment of the former would require to go beyond an effective LO+LL accuracy,
by combining NLO QCD corrections to the loop-induced gluon-fusion contribution with NNLL resummation. Given its small numerical impact, we leave such study for future work.

In order for the theoretical prediction to be reliable over the entire spectrum, the resummation of large logarithms at small \pt{} must be combined with the fixed-order cross section, valid at high \pt{}.
We consistently match N$^3$LL resummation for the \ptllg{} spectrum with NNLO corrections at the level of cumulative cross section, defined as ($\kappa$ = NNLO, N$^3$LL)
\begin{equation}
\label{eq:cumulative}
\sigma_{\kappa}(\ptveto) \equiv \int_0^{\ptveto} \rd \ptllg\; \frac{\rd \sigma_{\kappa}(\ptllg)}{\rd \ptllg}\,.
\end{equation}
The cross sections should be understood as being fully differential in the 
Born phase space, which allows us to apply arbitrary IR-safe cuts on the kinematics 
of the leptons and the photon.

There is a certain level of freedom when defining matching procedures that differ from one another only by terms
beyond the formal accuracy of the calculation. 
We study two different matching schemes.
The first scheme we consider is a customary additive scheme, which at NNLO+N$^3$LL is defined as
\begin{align}\label{eq:additive}
{\rm \sigma}^{\rm add.\, match.}_{\rm NNLO+N^3LL}(\ptveto)  = \sigma_{\rm NNLO}(\ptveto) - \left[\sigma_{\rm N^3LL}(\ptveto)\right]_{\rm NNLO} + \sigma_{\rm N^3LL }(\ptveto)\,.
\end{align} 
The notation $[\ldots]_{{\rm N}^k{\rm LO}}$ is used to indicate that the expression inside the bracket is expanded in $\as{}$ and truncated at N$^k$LO. Thus, the second term corresponds to the expansion of the resummed cumulative cross section $\sigma_{\rm N^3LL }(\ptveto)$ up to NNLO, i.e. ${\cal O}(\as^2)$, which subtracts all logarithmically enhanced contributions at small $\ptveto$ from the fixed-order component. This term is necessary to render \eqn{eq:additive} finite in the $\ptveto\to0$ limit and to remove the double counting between the 
first and the third term. 

The second scheme we consider is a multiplicative scheme~\cite{Bizon:2018foh,Bizon:2019zgf}, defined as
\begin{equation}
\label{eq:multiplicative1}
{\rm \sigma}^{\rm mult. \,match.}_{\rm NNLO+N^3LL}(\ptveto) = \frac{{\rm \sigma}_{\rm N^3LL}(\ptveto)}{{\rm \sigma}_{\rm N^3LL}^{\rm asym.} } \left[{\rm \sigma}_{\rm N^3LL}^{\rm asym.} \frac{{\rm \sigma}_{\rm NNLO}(\ptveto)}{\left[\sigma_{\rm N^3LL}(\ptveto)\right]_{\rm NNLO}}\right]_{\rm NNLO},
\end{equation}
where $ {\rm \sigma}_{\rm N^3LL}^{\rm asym.}$ is the asymptotic ($\ptveto \to\infty$) limit of the resummed cross section. 
In the limit $\ptveto \to 0$, \eqn{eq:multiplicative1} yields the resummed prediction, while for 
$\ptveto \to\infty$  it reproduces the fixed-order result.
The detailed matching formul\ae\ for the multiplicative scheme are reported in
appendix~A of ref.~\cite{Bizon:2018foh}.
Since in both matching schemes the cumulative cross section tends to $\sigma_{\rm NNLO}$ when $\ptveto\to\infty$, by construction the differential distribution fulfils the unitarity constraint, i.e.\ its integral yields the NNLO cross section.

We present predictions for the LHC at 13\,TeV. The EW parameters are evaluated 
through the $G_\mu$ scheme by setting the EW coupling to 
$\alpha=\sqrt{2}\,G_F m_W^2\left(1-m_W^2/m_Z^2\right)/\pi$ and the mixing angle
to $\cos\theta_W^2=(m_W^2-i\Gamma_W\,m_W)/(m_Z^2-i\Gamma_Z\,m_Z)$, employing
the complex-mass scheme~\cite{Denner:2005fg} throughout.
We choose the PDG~\cite{Patrignani:2016xqp} values for the the input parameters:
$G_F = 1.16639\times 10^{-5}$\,GeV$^{-2}$, $m_W=80.385$\,GeV,
$\Gamma_W=2.0854$\,GeV, $m_Z = 91.1876$\,GeV, $\Gamma_Z=2.4952$\,GeV.
For each perturbative order we use the corresponding set of $N_f=5$ 
NNPDF3.0~\cite{Ball:2014uwa} parton distributions with $\as(m_Z)=0.118$.
The renormalization scale ($\mu_R$) and the factorization scale ($\mu_F$) are chosen dynamically as
\begin{align} 
\mu_R=\mu_F=\mu_0\equiv\,\sqrt{m_{\ell\ell}^2+\ptg^2}\,,
\end{align}
while the resummation scale ($Q$) is set to
\begin{align}
Q=Q_0\equiv \frac12 \, m_{\ell\ell\gamma}\,.
\end{align}
Uncertainties from missing higher-order contributions 
are estimated from customary 7-point renormalization- and factorization-scale variations by a factor of two around $\mu_0$ for $Q=Q_0$ 
with the constraint $0.5\le \mu_R/\mu_F\le 2$, and by varying $Q$ by a factor of two around $Q_0$ for $\mu_F = \mu_R = \mu_0$.
The total scale uncertainty is evaluated as the envelope of the resulting nine variations.
The resummation is turned off at high \ptllg{} by means of modified logarithms as defined in \citere{Bizon:2017rah}, with exponent $p=4$. We have 
checked that our predictions have a negligible dependence on the value of $p$.
Non-perturbative corrections have not been included in our results.

We study predictions for the \ptllg{} distribution 
in two setups that involve different phase-space selection cuts, defined 
in \tab{tab:cuts}: The first is a loose selection that solely aims at preventing QED singularities, including transverse-momentum and rapidity requirements for the photon, a lower invariant mass cut on the 
lepton--photon system, a $Z$-mass window for the lepton pair, and Frixione smooth-cone isolation \cite{Frixione:1998jh}. This 
setup will be referred to as ``inclusive" in the following.
The second setup corresponds to the fiducial selection of the 13\,TeV ATLAS analysis of \citere{Aad:2019gpq},
which uses a tighter requirement for the transverse momentum of the photon, 
a lower invariant-mass cut on the lepton pair, transverse-momentum and rapidity requirements 
on the leading and subleading lepton, a lower bound for the sum of the invariant masses of the lepton pair and the $\ell\ell\gamma$
system, a lepton--photon separation in $\Delta R=\sqrt{\Delta \phi^2+\Delta\eta^2}$, and a two-fold photon isolation: 
in addition to a Frixione isolation with a rather small cone, the transverse energy of hadrons collimated with the photon is required not to exceed a small fraction of its transverse momentum.
In our parton-level calculation we define $p_T^{\rm cone0.2}$ as the sum of the transverse momenta 
of all partons within a cone of $R=0.2$ around the photon. 
The second setup is referred to as ``fiducial" in the following.
One should bear in mind that such isolation criteria induce NG logarithmic corrections, which we do not resum in our formalism.
We will estimate their effect on the \ptllg{}  spectrum below.

\renewcommand{\baselinestretch}{1.5}
\begin{table}[!t]
\begin{center}
\begin{tabular}{c}
\toprule
inclusive setup for $pp\to\ell\ell'\gamma +X,\quad \ell,\ell'\in\{e,\mu\}$\\
\midrule
$\ptg\ge10$\,GeV,\quad $|\eta_\gamma|\le 2.37$,\quad $m_{\ell\gamma}\ge 4$\,GeV, \quad$66\,\textrm{GeV}\le m_{\ell\ell} \le 116$\,GeV,\\
Frixione isolation with $n=2$, $\delta_0=0.1$, and $\epsilon = 0.1$\,.\\
%$\Delta R_{\ell\ell} >0.2$, \quad$\Delta R_{\ell\ell'} >0.2$\\
%anti-$k_T$ jets with $R=0.4$, $p_{T,j}>25$\,GeV, $|\eta_j|<4.5$\\
%lepton identification in SF channel:\\[-0.2cm]
%minimizing differences of invariant-mass of OSSF lepton pairs and $m_Z$\\
\bottomrule
fiducial setup for $pp\to\ell\ell'\gamma +X,\quad \ell,\ell'\in\{e,\mu\}$; used in the ATLAS 13\,TeV analysis of~\citere{Aad:2019gpq}\\
\midrule
$\ptlone\ge 30$\,GeV, \quad $\ptltwo\ge 25$\,GeV, \quad$|\etal|\le 2.47$,\quad $m_{\ell\ell}\ge 40$\,GeV,\\
$\ptg\ge 30$\,GeV,\quad $|\eta_\gamma|\le 2.37$,\quad $m_{\ell\ell}+m_{\ell\ell\gamma}\ge 182$\,GeV,\quad$\Delta R_{\ell\gamma} >0.4$,\\
Frixione isolation with $n=2$, $\delta_0=0.1$, and $\epsilon = 0.1$\,,\quad $p_T^{\rm cone0.2}/\ptg< 0.07$.\\
%
%Axial-$\ptmiss>90$\,GeV, \quad$\pt\textrm{-balance}<0.4$,\\
%$\njets=0$, \quad anti-$k_T$ %\cite{}
%jets with $R=0.4$, $p_{T,j}>25$\,GeV, $|\eta_j|<4.5$ and $\Delta R_{ej}> 0.3$\\
\bottomrule
\end{tabular}
\end{center}
\renewcommand{\baselinestretch}{1.0}
\caption{\label{tab:cuts} %Definition of the fiducial volumes 
Definition of phase-space cuts.}
\vspace{-0.5cm}
\end{table}
\renewcommand{\baselinestretch}{1.0}

%
%expansion vs fixed order NLO vs NLLexp and NNLO vs N3LLexp; inclusive+fiducial
%
%inclusive: left NLL, NLO, NLO+NNLL; right: N3LL, NNLO, NNLO+N3LL
%inclusive: scale variation muR muF only; scale variation Q only
%
%inclusive left, fiducial right: NNLO+N3LL vs NLO+NLL
%inclusive left, fiducial right: NNLO+N3LL additive vs multiplicative
%
%NNLO+N3LL best vs ATLAS data

\begin{figure}
\begin{center}
\begin{tabular}{cc}
\includegraphics[width=.33\textheight]{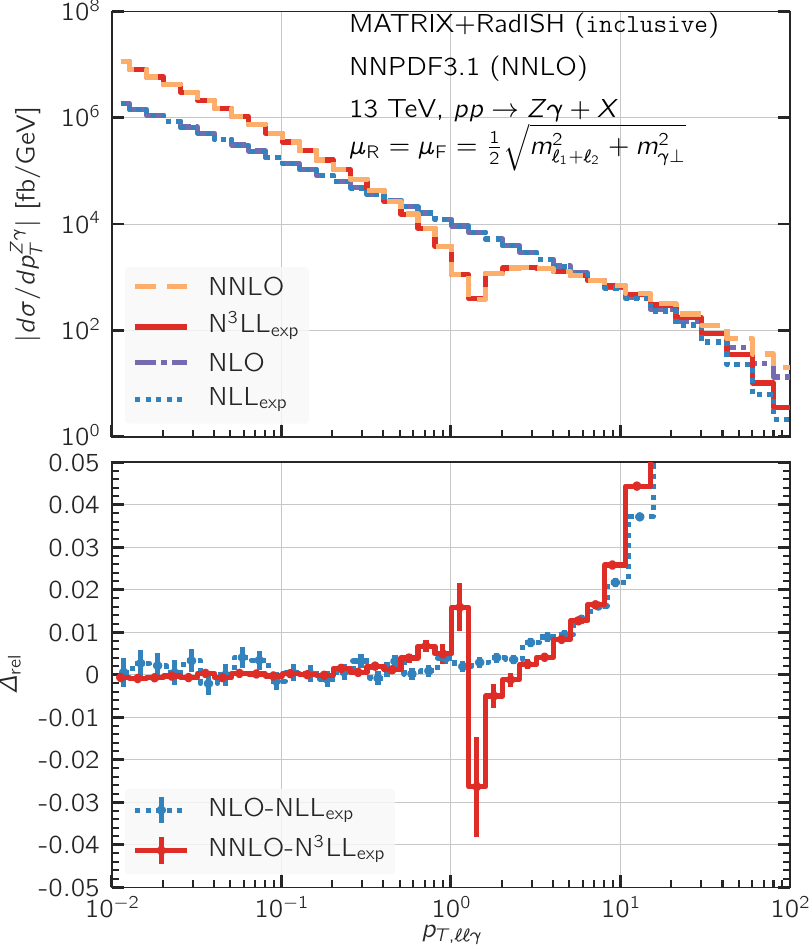} 
&
\includegraphics[width=.33\textheight]{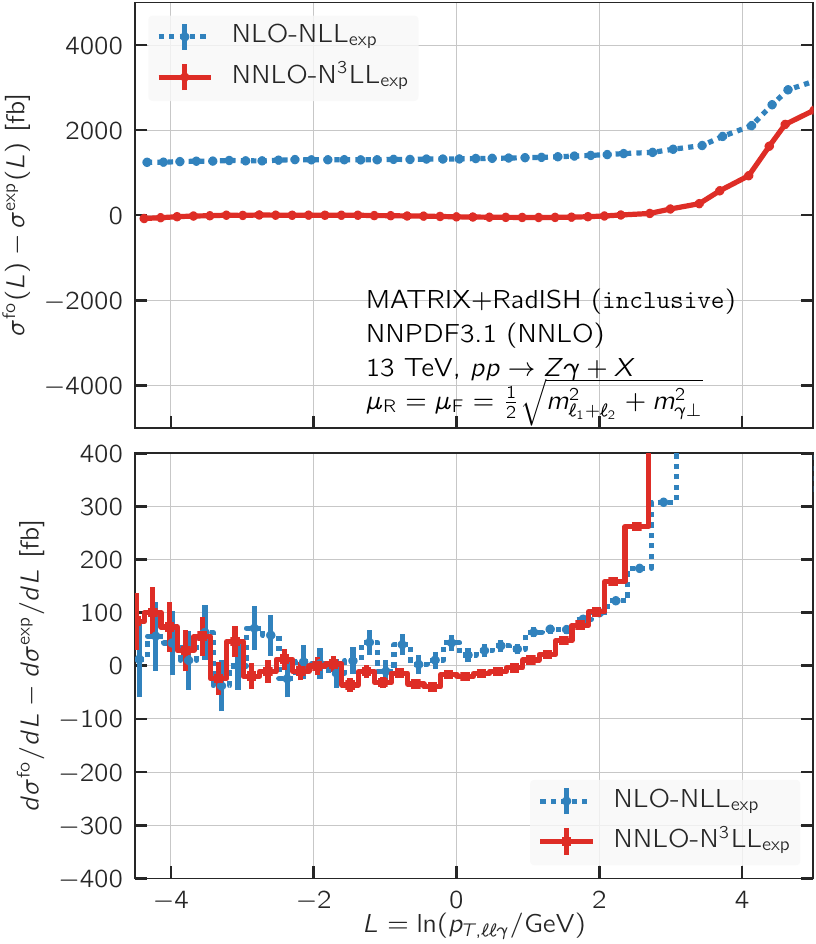}
 \\
\hspace{0.6em} (a) & \hspace{1em}(b)
\end{tabular}\vspace{0.2cm}
\begin{tabular}{cc}
\includegraphics[width=.33\textheight]{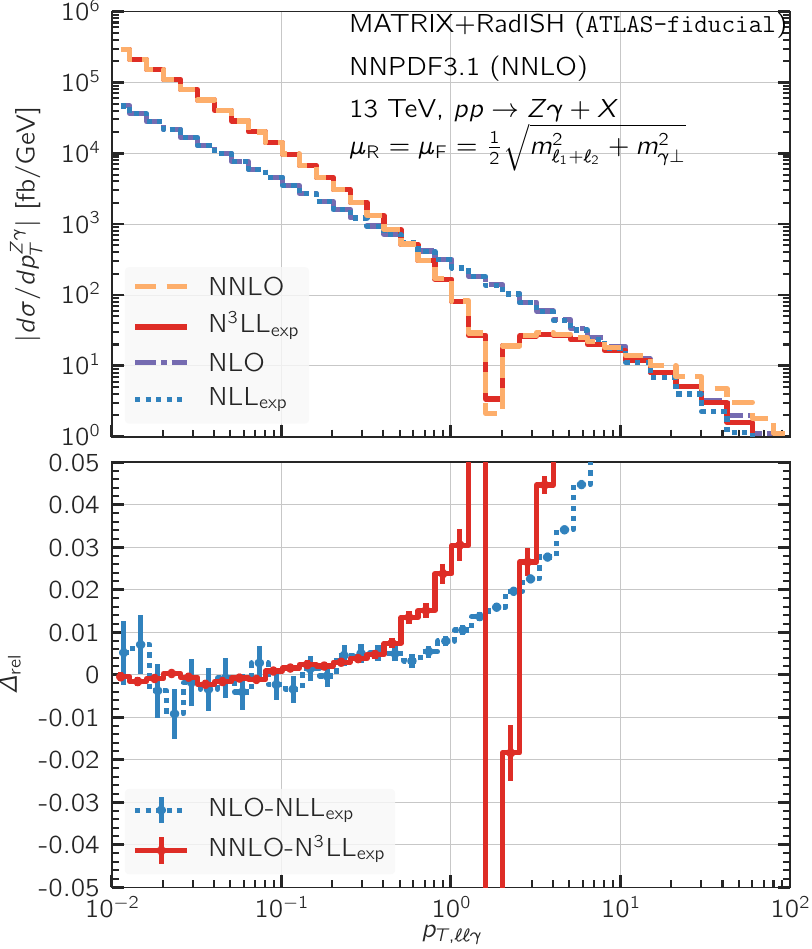}
&
\includegraphics[width=.33\textheight]{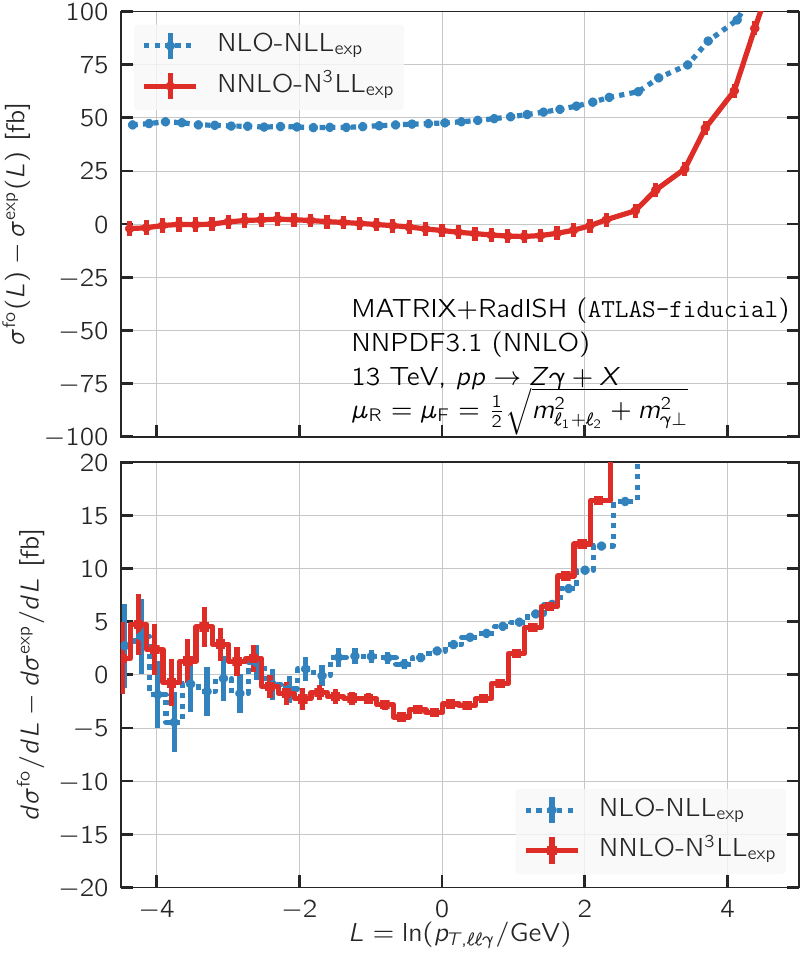}\\
\hspace{0.6em} (c) & \hspace{1em}(d)
\end{tabular}\vspace{0.5cm}
  \caption{\label{fig:validation} Panel (a) and (c): transverse-momentum spectrum of the $Z\gamma$ pair 
  at NLO (purple, dot-dashed) and NNLO (orange, dashed), and the expansion of the NLL (blue, dotted) and N$^3$LL (red, solid) cross section. 
  The lower frame shows the relative difference between the fixed-order cross section and the expansion, normalized to the latter.
  Panel (b) and (d): the upper frame shows the difference at the cumulative level between NLO and NLL expansion (blue, dotted), and between NNLO and N$^3$LL expansion (red, solid). The lower frame shows the same results for the derivative of the cumulative cross 
  section with respect to $\ln(\ptllg/{\rm GeV})$.}
\end{center}
\end{figure}

We start the discussion of our results by comparing the expansion of the resummation with the fixed-order spectrum at small \ptllg{} in \fig{fig:validation}, which provides a strong check of our calculation. The plots demonstrate at a remarkable precision that
the expansion of the resummed cross section matches the fixed-order cross section 
at small transverse momenta both for the inclusive setup in panel (a) and (b)  and the fiducial setup in panel (c) and (d). 
As can be seen from the lower frame in panel (a) and (c), the relative difference $\Delta_{\rm rel}$ between the 
NNLO distribution and the NNLO expansion of the N$^3$LL distribution normalized to the latter 
(red solid curve) vanishes down to $\ptllg=0.01$\,GeV within the numerical errors at the permille level.
In the upper frame of panel (b) and (d) we show the difference of the 
NNLO cross section and NNLO expansion of the N$^3$LL cross section at the cumulative level (red solid curve).
Since their difference tends to zero at low transverse momenta, also constant terms in \ptllg{} match 
between NNLO and N$^3$LL.
The fact that at NLO the difference with the NLL expansion (blue dotted curve) tends to a constant different from zero is expected,
since our NLL result does not include the constant NLO terms in \ptllg{}.
Finally, the lower frame in panel (b) and (d) shows that the absolute difference 
between the fixed-order result and the expansion of the resummation after taking the derivative 
of the cumulative cross sections with respect to $\ln(\ptllg/{\rm GeV})$ yields zero within numerical uncertainties at small transverse momenta.
This indicates that all logarithmic terms in \ptllg{} are correctly predicted.
Not only do these comparisons provide stringent checks of the validity of our calculation, but they also 
show the excellent precision that our numerical framework can achieve.

We further notice from these plots that the logarithmically enhanced contributions become dominant over the regular 
terms at smaller values of transverse momentum compared to other processes (cf.~\citere{Kallweit:2020gva} for instance).
Especially in the fiducial setup, regular contributions become non-negligible already at $\ptllg\sim 1\,$\,GeV. Indeed, it has 
been shown before \cite{Grazzini:2017mhc,Ebert:2019zkb} that processes with identified photons in the final state receive rather large corrections from 
power-suppressed terms at small transverse momentum.
In the multiplicative scheme \eqn{eq:multiplicative1} those are suppressed by ${\rm \sigma}_{\rm N^3LL}(\ptveto)$ at small \ptllg{}. 
Although such effects are beyond the nominal accuracy, this suppression may induce numerically relevant corrections, in particular in the fiducial setup considered here. This behaviour is undesirable, since these power corrections are a genuine non-singular contribution to the cross section.
For this reason a multiplicative scheme is not ideal
when the fixed-order cross section features large power-suppressed corrections, 
and we choose the additive scheme as the default throughout this paper.\footnote{We stress that the multiplicative scheme, which is the default in {\sc Matrix+RadISH}, has advantages in cases where power corrections are moderate. In particular it is numerically more stable at small transverse momenta, and it 
includes the constant contributions in \pt{} through the matching when those
are not available in the resummation component.}

\begin{figure}
\begin{center}
\begin{tabular}{cc}
\includegraphics[width=.33\textheight]{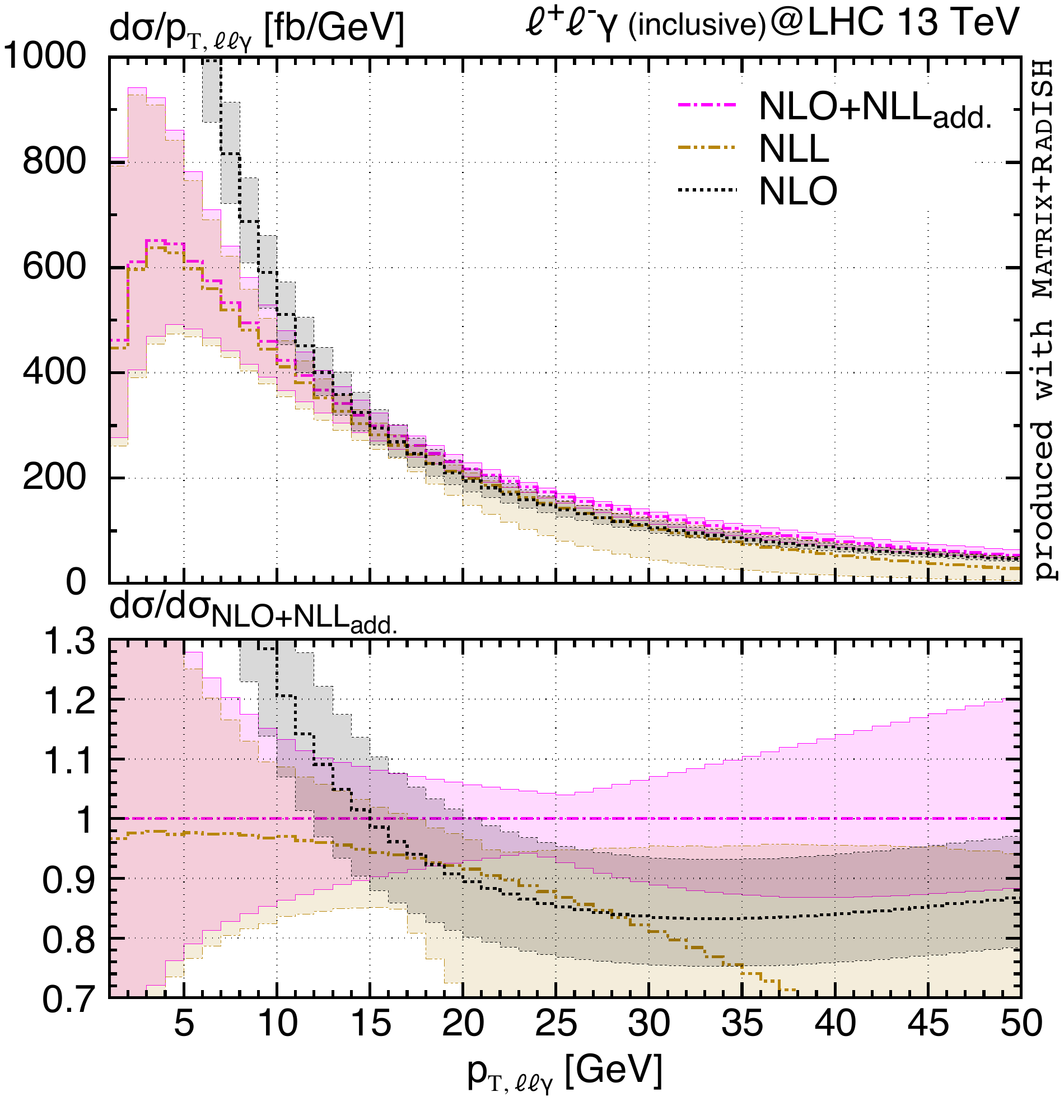} 
&
\includegraphics[width=.33\textheight]{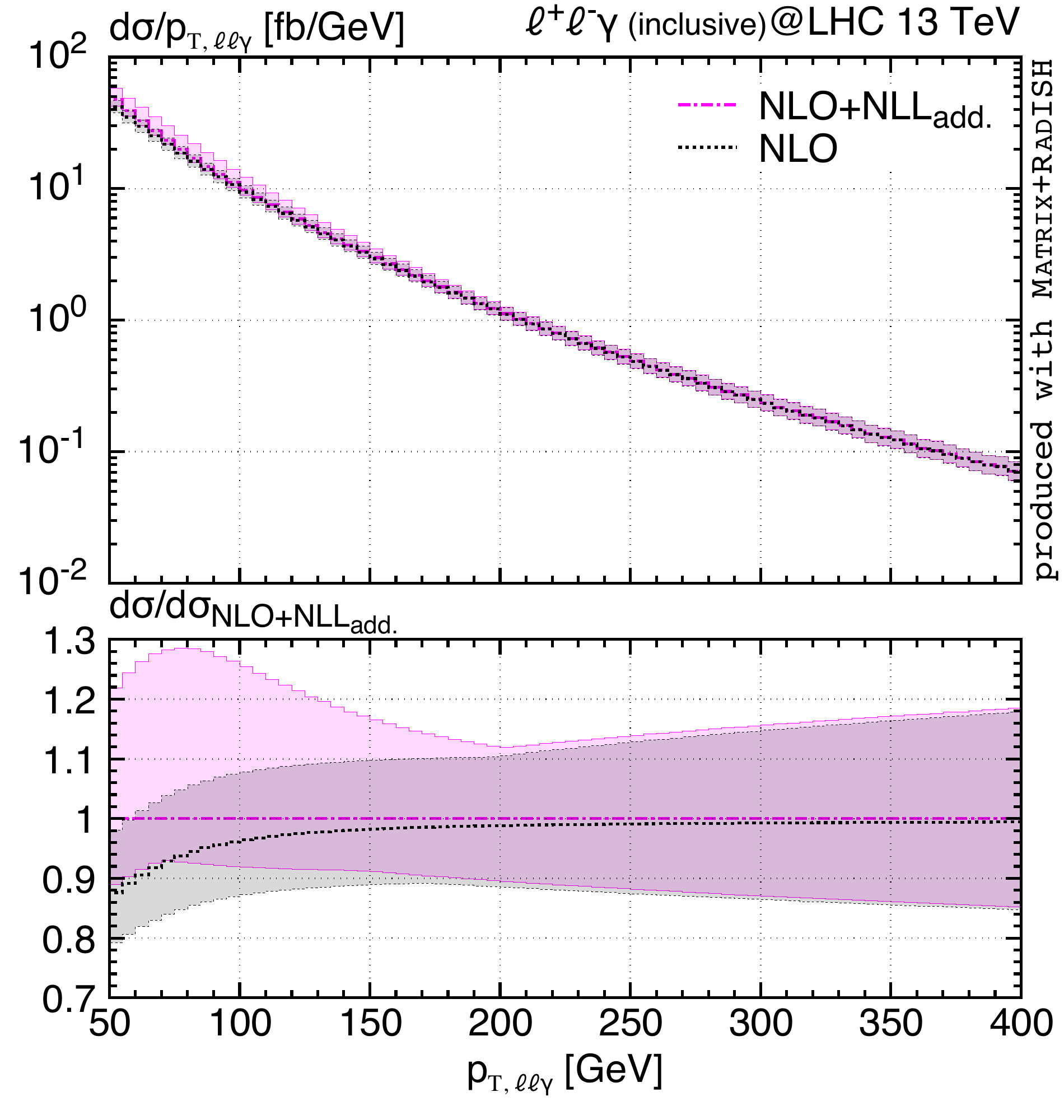}
 \\
\hspace{0.6em} (a) & \hspace{1em}(b)
\end{tabular}\vspace{0.5cm}
\begin{tabular}{cc}
\includegraphics[width=.33\textheight]{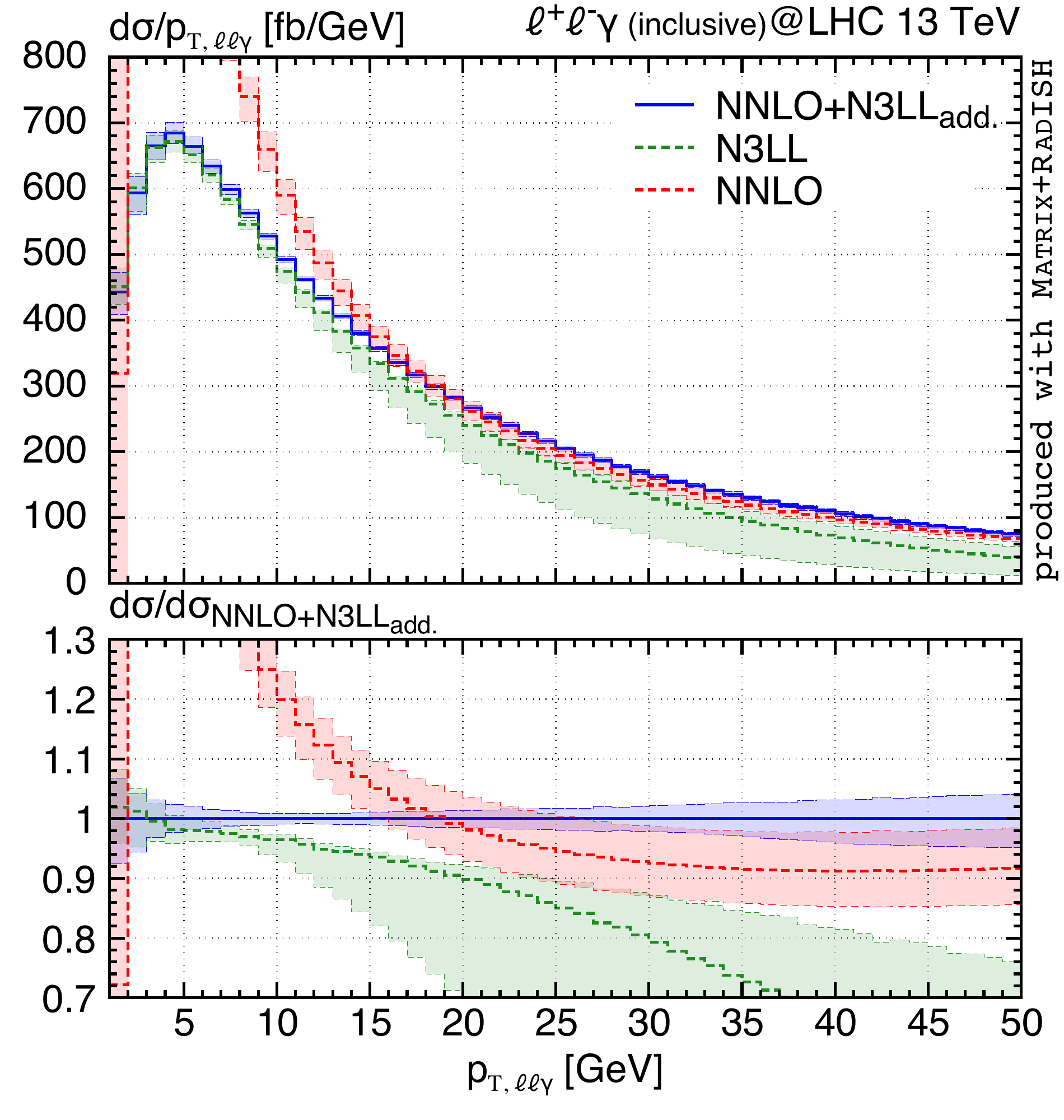}
&
\includegraphics[width=.33\textheight]{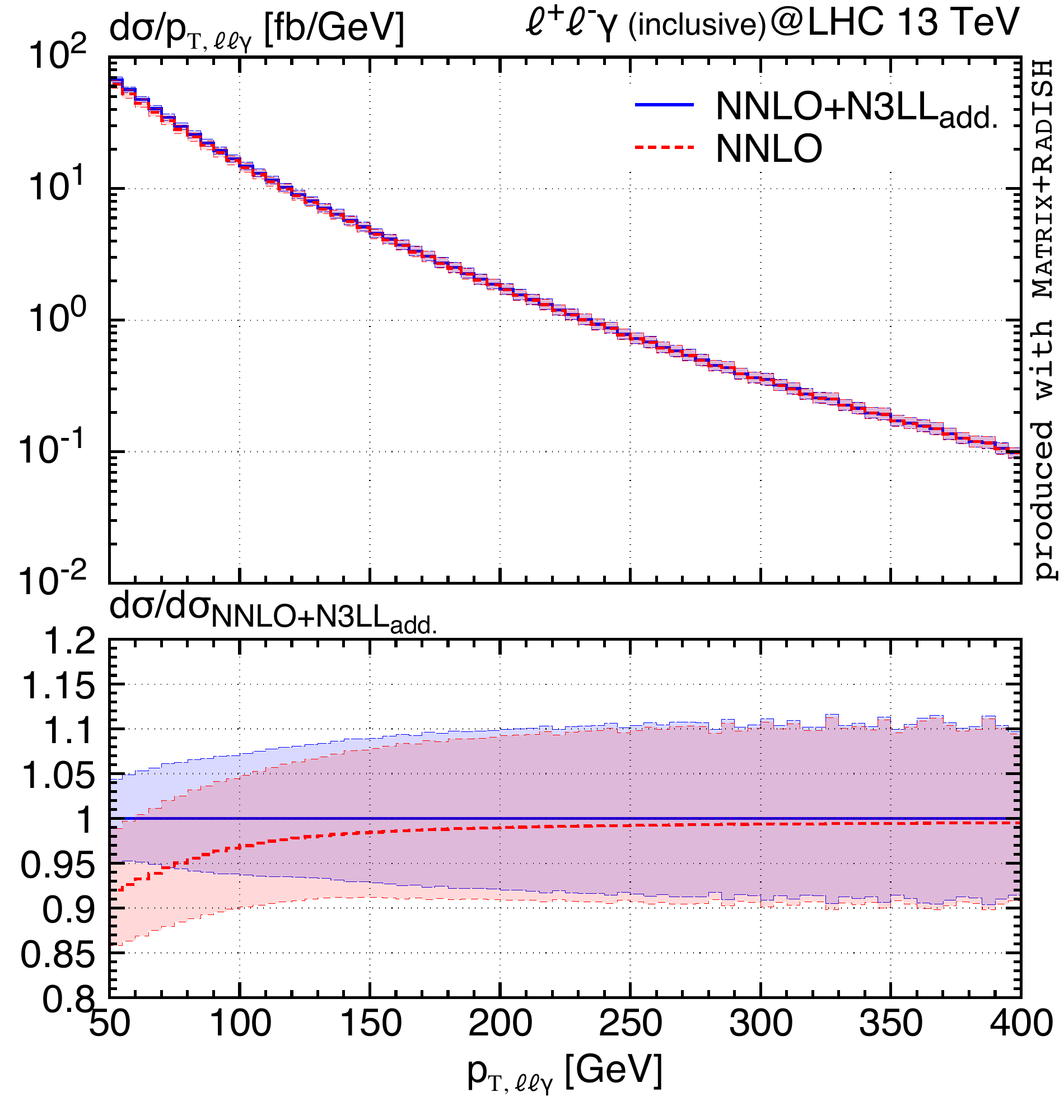}\\
\hspace{0.6em} (c) & \hspace{1em}(d)
\end{tabular}\vspace{0.5cm}
  \caption{\label{fig:inclusive} Panel (a) and (b): \ptllg{} spectrum at NLO (black, dotted), NLL (brown, dash-double-dotted), and NLO+NLL (magenta, dash-dotted) 
  in the small-\ptllg{} (panel (a)) and large-\ptllg{} (panel (b)) region. The lower frames show the ratio to the central NLO+NLL prediction. 
 Panel (c) and (d): \ptllg{} spectrum at NNLO (red, dashed), N$^3$LL (green, double-dash-dotted), and NNLO+N$^3$LL (blue, solid) 
  in the small-\ptllg{} (panel (a)) and large-\ptllg{} (panel (b)) region. The lower frames show the ratio to the central NNLO+N$^3$LL prediction.}
\end{center}
\end{figure}

We now turn to discussing the resummed transverse-momentum spectrum of the $Z\gamma$ pair. \fig{fig:inclusive} shows results 
in the inclusive setup and compares the matched NLO+NLL spectrum to the NLL and the NLO results in panel (a) and (b), and 
the matched NNLO+N$^3$LL spectrum to the N$^3$LL and the NNLO results in panel (c) and (d). At large transverse momenta (panel (b) and (d)),
the matched results nicely converge towards the fixed-order predictions. 
At small transverse momenta (panel (a) and (c)), the NLO and NNLO predictions become unreliable, while the resummation yields physical results. 
The matched predictions are very close to the purely resummed ones at small transverse momenta and then progressively move 
farther apart at larger \ptllg{}. Looking at the scale uncertainties, we observe a substantial 
reduction in the size of the respective bands when moving from NLO+NLL to NNLO+N$^3$LL: At large \ptllg{} they decrease by roughly a 
factor of two, from $20$\% to $10$\%. At small \ptllg{} the reduction is even more significant. For $\ptllg\lesssim 20$\,GeV the
NLO+NLL uncertainty increases between about $10$\% to more than $30$\%, while it is at the few-percent level at NNLO+N$^3$LL, reaching 
at most $\sim 8$\% in the first bin.

\begin{figure}
\begin{center}
\begin{tabular}{cc}
\includegraphics[width=.33\textheight]{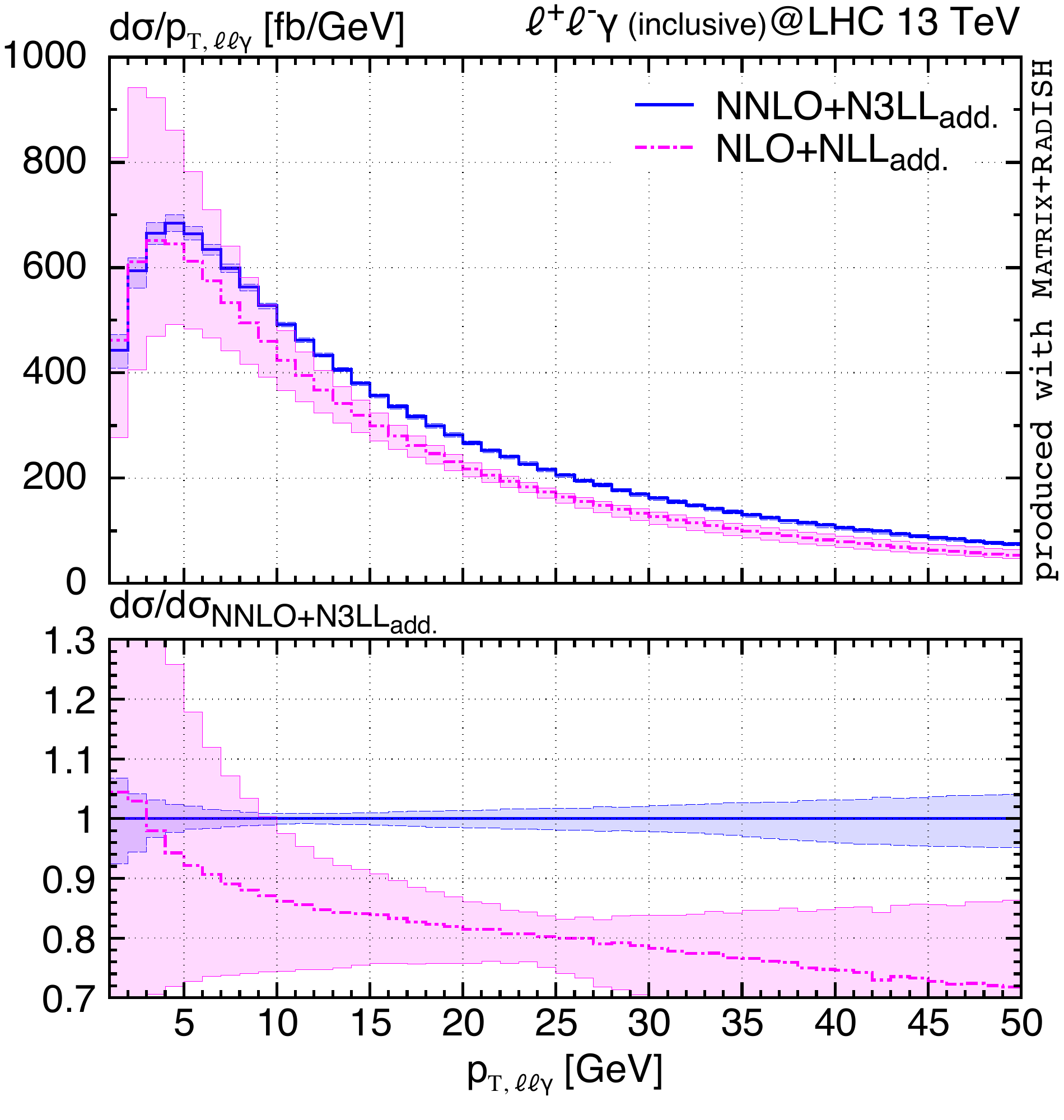} 
&
\includegraphics[width=.33\textheight]{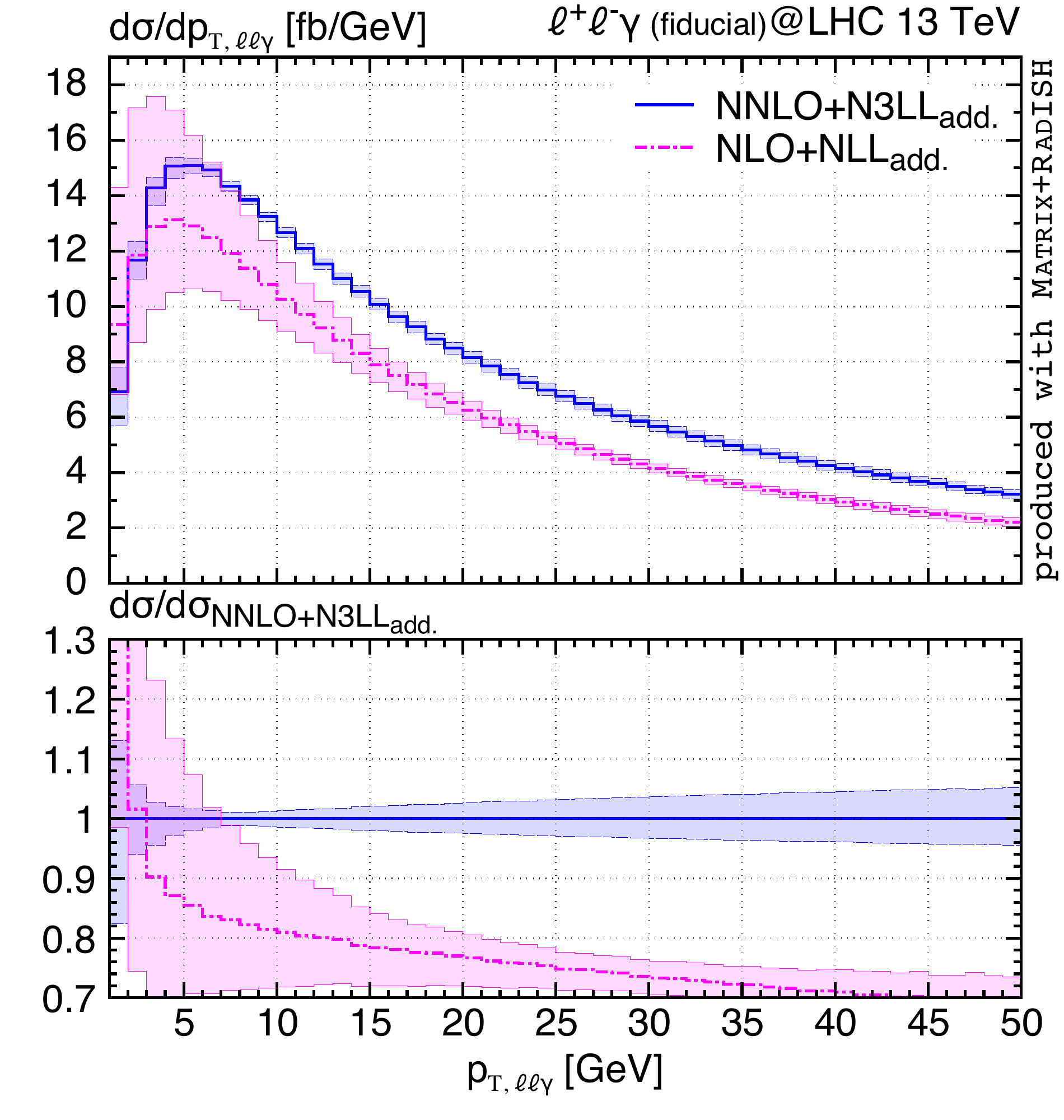}
 \\
\hspace{0.6em} (a) & \hspace{1em}(b)
\end{tabular}\vspace{0.5cm}
  \caption{\label{fig:NvsNN} \ptllg{} spectrum at NLO+NLL (magenta, dash-dotted) and NNLO+N$^3$LL (blue, solid) 
  in the inclusive (panel (a)) and fiducial (panel (b)) setup. The lower frames show the ratio to the central NNLO+N$^3$LL prediction.}
\end{center}
\end{figure}

\fig{fig:NvsNN} compares directly the results at NLO+NLL and NNLO+N$^3$LL at small transverse momenta, both in the inclusive setup in panel (a) and 
in the fiducial setup in panel (b). Higher-order corrections move the peak by $1$--$2$\,GeV towards larger values of \ptllg{}.
The substantial reduction of scale uncertainties has already been pointed out for the inclusive case, and we find 
a quite similar picture in the fiducial case. For $\ptllg\lesssim 10$\,GeV NLO+NLL and NNLO+N$^3$LL results agree within uncertainties with each other,
although the corrections at $\ptllg= 10$\,GeV are already about $15$\% in both setups. With increasing values of \ptllg{} the corrections become 
progressively larger, reaching about $30$\% at $\ptllg= 50$\,GeV. For $\ptllg\gtrsim 10$\,GeV the higher-order corrections are not covered by the 
scale-uncertainty band of the NLO+NLL prediction, which appears to be significantly underestimated. This behaviour is not unexpected, as it is 
directly inherited from the fixed-order calculation, where the relatively small NLO uncertainties do not cover the substantial NNLO corrections 
in the tail. We stress that in the tail of the \ptllg{} distribution the (N)NLO prediction is effectively only (N)LO accurate, which explains 
this observation, as LO uncertainties generally tend to underestimate higher-order effects. While we find a fairly similar pattern in the two setups,
the additional fiducial cuts tend to slightly increase the relative size of the corrections.

\begin{figure}
\begin{center}
\begin{tabular}{cc}
\includegraphics[width=.33\textheight]{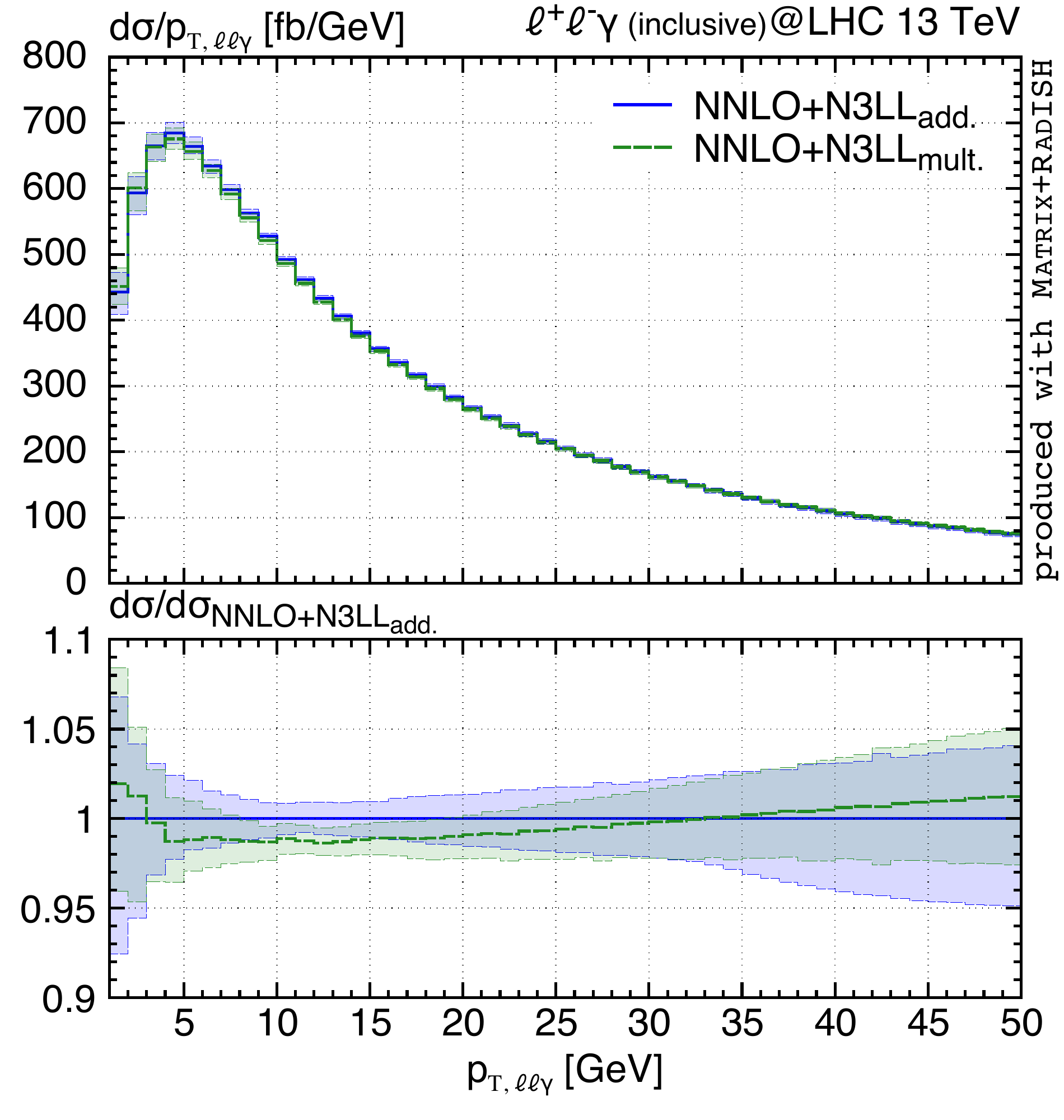} 
&
\includegraphics[width=.33\textheight]{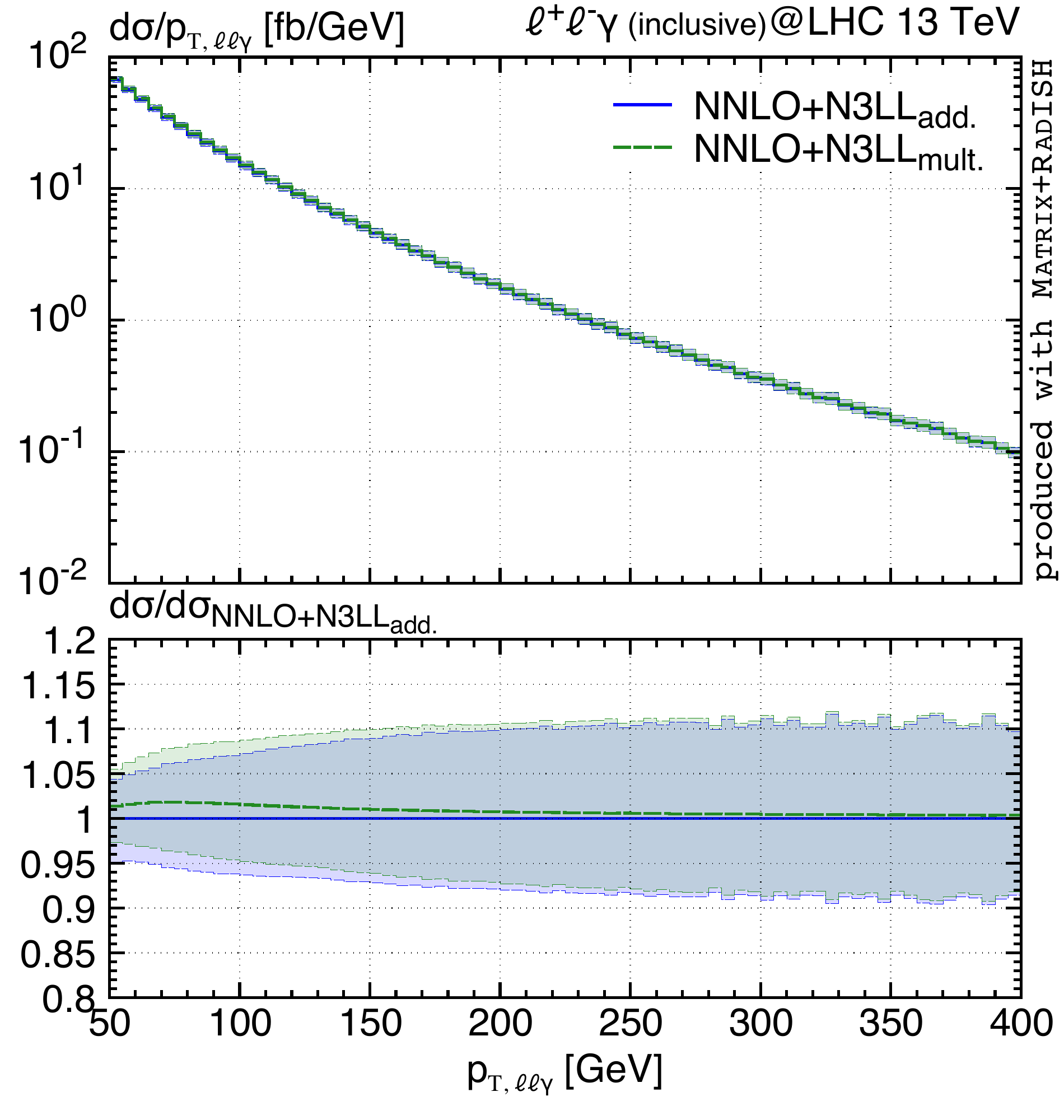}
 \\
\hspace{0.6em} (a) & \hspace{1em}(b)
\end{tabular}\vspace{0.5cm}
\begin{tabular}{cc}
\includegraphics[width=.33\textheight]{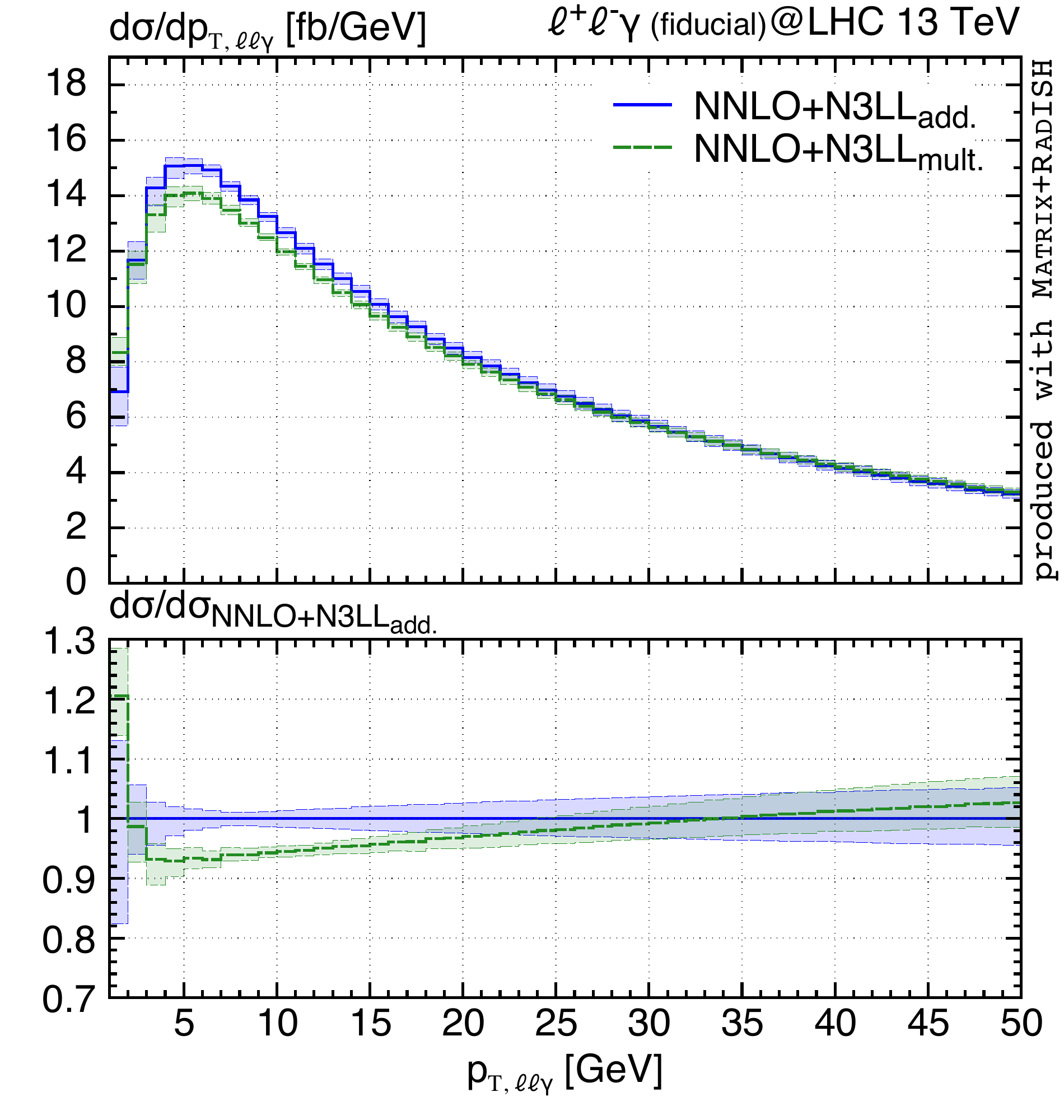}
&
\includegraphics[width=.33\textheight]{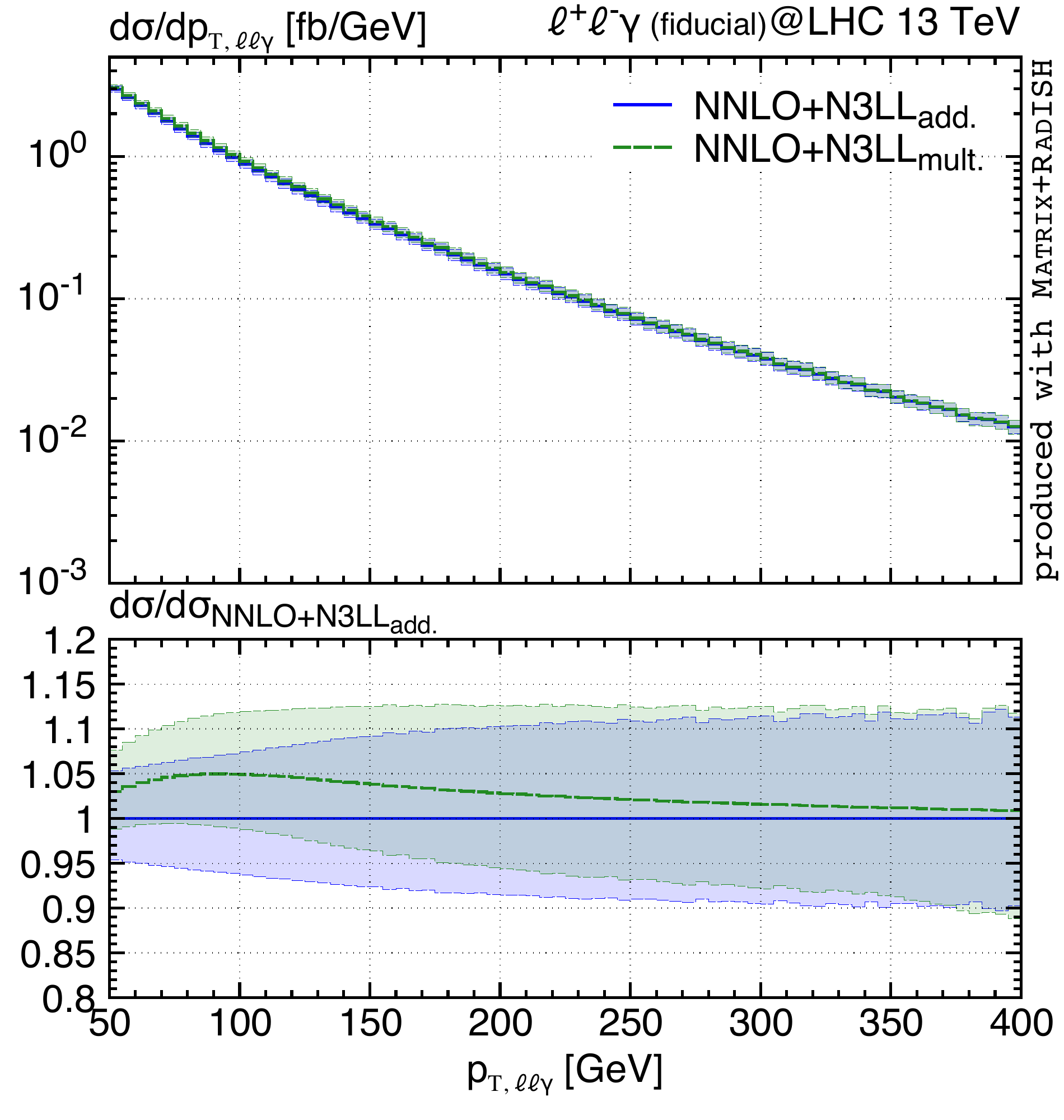}\\
\hspace{0.6em} (c) & \hspace{1em}(d)
\end{tabular}\vspace{0.5cm}
  \caption{\label{fig:addvsmult} \ptllg{} spectrum at NNLO+N$^3$LL in the additive matching scheme (blue, solid) and in the multiplicative
  matching scheme (green, long-dashed) in the inclusive (panel (a) and (b)) and fiducial (panel (c) and (d)) setup, showing the small-\ptllg{} (panel (a) and (c)) and large-\ptllg{} (panel (b) and (d)) region. 
  The lower frames show the ratio to the central prediction in the additive scheme.}
\end{center}
\end{figure}

In \fig{fig:addvsmult} we compare our default NNLO+N$^3$LL predictions in the additive matching scheme of \eqn{eq:additive} with NNLO+N$^3$LL predictions in the multiplicative matching scheme of \eqn{eq:multiplicative1}
for the inclusive case in panel (a) and (b), and for the fiducial case in panel (c) and (d). For large transverse momenta (panel (b) and (d)), the two predictions are in good agreement within uncertainties, 
since both eventually approach the NNLO result in the tail of the distribution. At small transverse momenta (panel (a) and (c)), the situation is different for the two setups. By and large, in the inclusive setup
we find good agreement between the two matching schemes with overlapping uncertainty bands and at most $2$\% differences in the central value. The difference can be
understood as an uncertainty related to the inclusion of terms beyond nominal accuracy in the matched prediction. In the fiducial setup the differences are somewhat larger. Multiplicative and additive schemes 
differ by up to $\sim 8$\% for \ptllg{} between $4$\,GeV and $20$\,GeV, and there is a gap between their uncertainty bands. This is in line with the large power
corrections observed in the fiducial setup in \fig{fig:validation}\,(c), which are 
suppressed in the multiplicative scheme and preserved in the additive one.
 As already stressed above, the suppression of such genuine non-singular contributions is undesirable, which justifies our preference for the additive scheme, especially in the fiducial setup, and we refrain from using the matching systematics as an additional uncertainty.

\begin{figure}[t]
\begin{center}
\begin{tabular}{cc}
\includegraphics[width=.33\textheight]{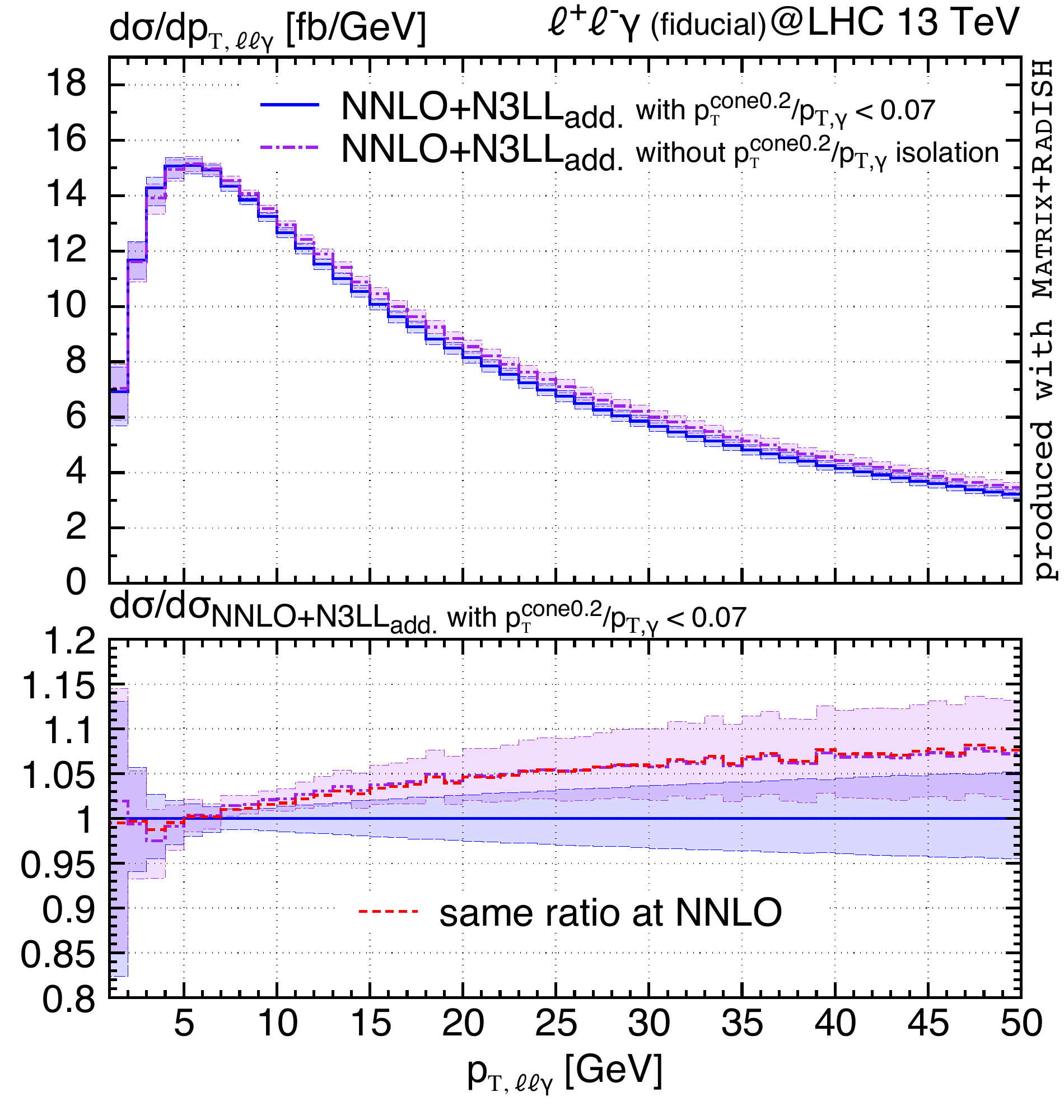} 
&
\includegraphics[width=.33\textheight]{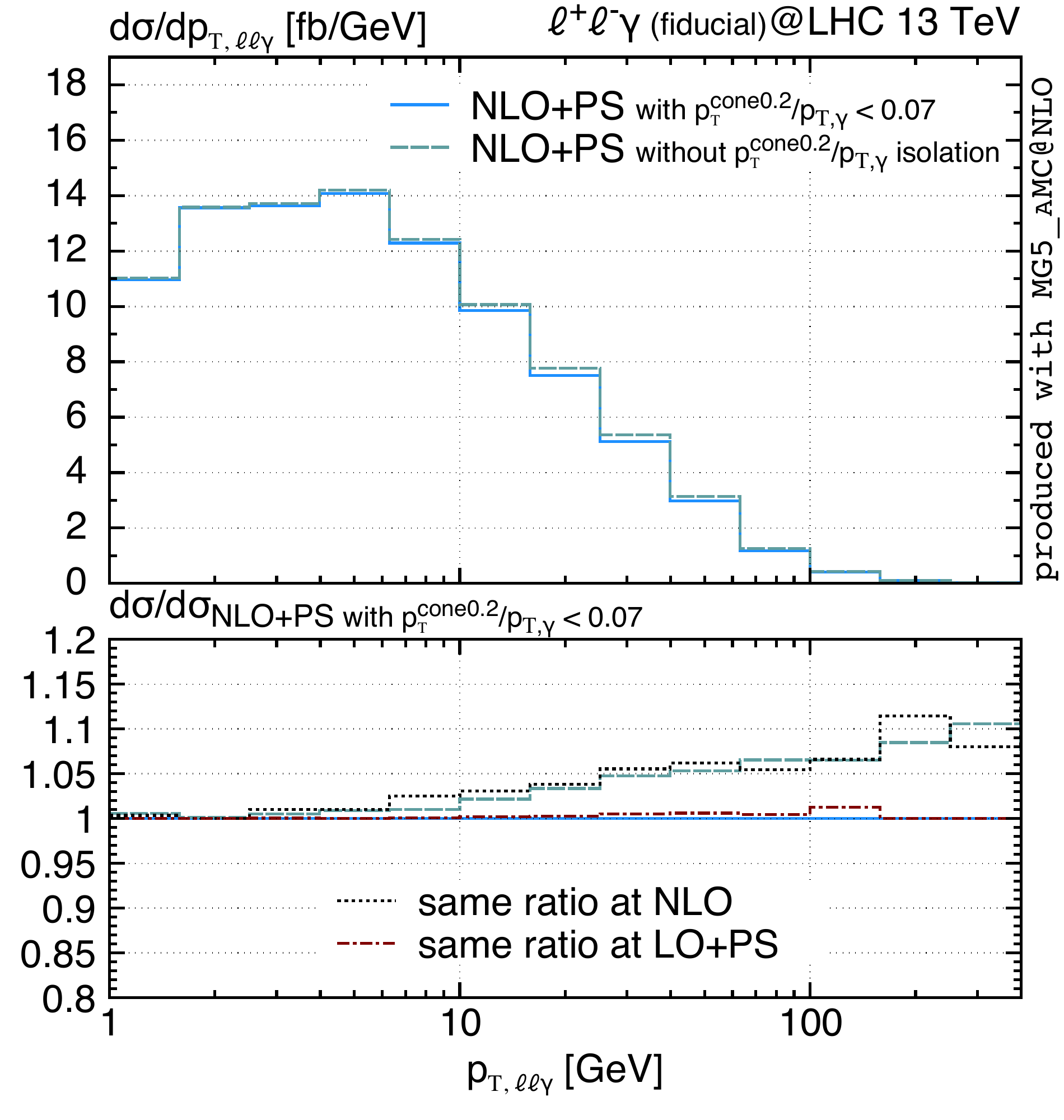}
 \\
\hspace{0.6em} (a) & \hspace{1em}(b)
\end{tabular}\vspace{0.5cm}
  \caption{\label{fig:nocone} Panel (a): \ptllg{} spectrum at NNLO+N$^3$LL in the fiducial setup (blue, solid) and without the additional $p_T^{\rm cone0.2}/\ptg$ isolation (purple, dash-dotted). The lower frame shows the ratio to the central prediction with the $p_T^{\rm cone0.2}/\ptg< 0.07$ requirement as well as the result when taking the same 
ratio for the central predictions at NNLO (red, dashed).
  Panel (b): Same results for central predictions at NLO+PS with $p_T^{\rm cone0.2}/\ptg< 0.07$ requirement (light blue, solid) and without (grey-blue, long-dashed). The lower frame shows the ratio to the 
former as well as the same ratios at NLO (black, dotted) and at LO+PS (brown, dash-dotted).}
\end{center}
\end{figure}

\begin{figure}[t]
\begin{center}
\begin{tabular}{cc}
\includegraphics[width=0.6\textwidth]{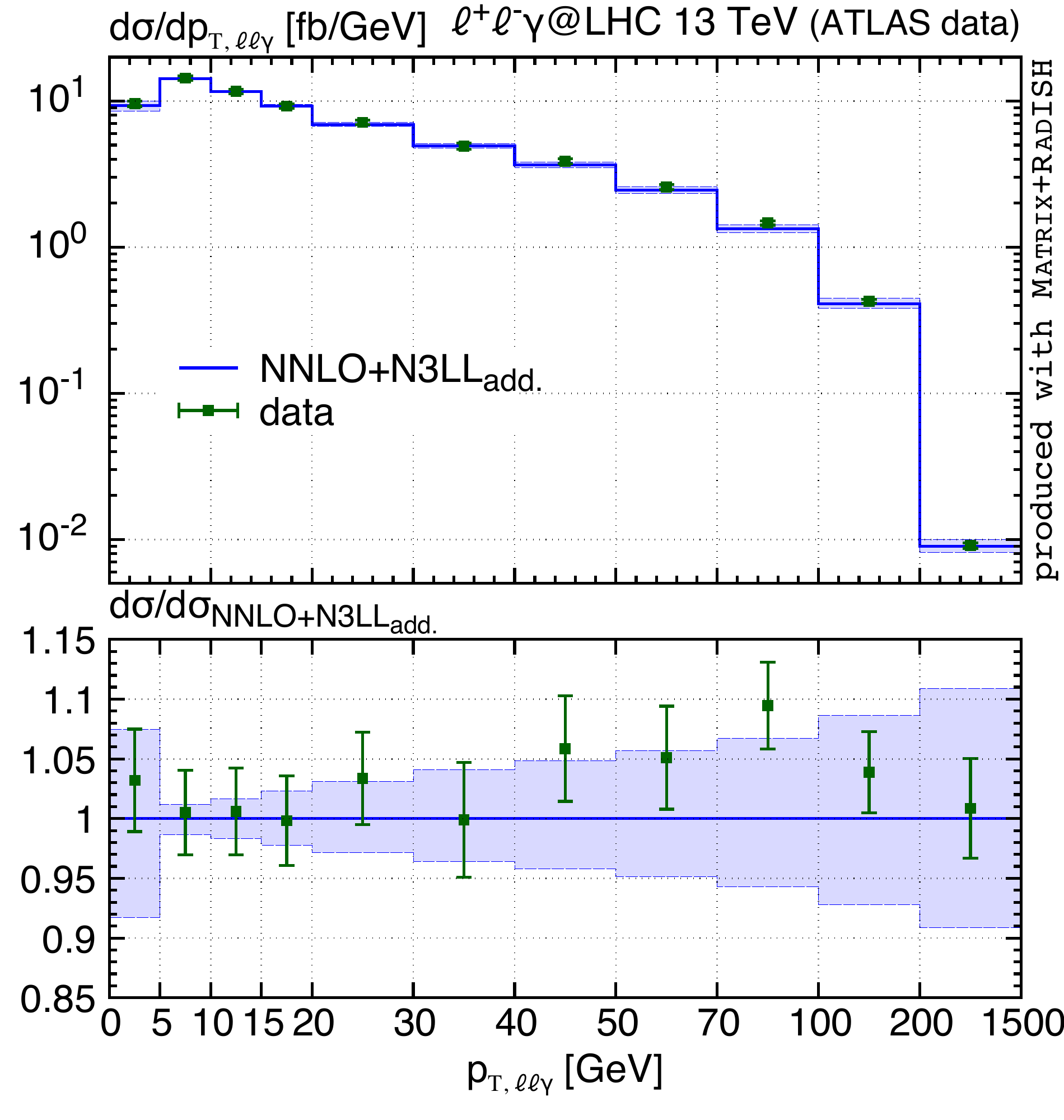}
\end{tabular}\vspace{0.5cm}
  \caption{\label{fig:atlas} NNLO+N$^3$LL prediction of the \ptllg{} spectrum (blue, solid) compared to ATLAS data \cite{Aad:2019gpq} (green data points). The lower frame shows the ratio to the central NNLO+N$^3$LL prediction.}
\end{center}
\end{figure}

We continue by studying the impact of NG logarithmic terms stemming from photon isolation in \fig{fig:nocone}.  Such terms are 
not included in our resummation approach and enter  
only through the matching to fixed order. \fig{fig:nocone}\,(a) compares
the NNLO+N$^3$LL predictions in the fiducial
setup with and without the $p_T^{\rm cone0.2}/\ptg< 0.07$ isolation cut
(which is additional to the smooth-cone isolation).
Their ratio indicates that at \ptllg{} values around the peak and smaller, 
the additional
isolation has a minimal impact, as expected being it power suppressed, while it induces effects 
of $\mathcal{O}(10\%)$ in the tail of the distribution. We show the 
same ratio at NNLO in the lower frame, which is essentially indistinguishable 
from the one at NNLO+N$^3$LL.
In other words, isolation effects are adopted purely from the fixed-order prediction. 
In fact, the small effect at
$\ptllg{} \lesssim 10$\,GeV indicates that the resummation of those corrections 
should have a minor impact in 
that region.
Furthermore, we estimate the all-order effects of including NG logarithmic contributions in the fiducial setup using the {\sc Pythia8}~\cite{Sjostrand:2014zea} parton shower (PS)
matched to NLO calculations in the MC@NLO scheme \cite{Frixione:2002ik} within {\sc MadGraph5\_aMC@NLO} \cite{Alwall:2014hca}. 
To this end, \fig{fig:nocone}\,(b) shows NLO+PS results with and 
without $p_T^{\rm cone0.2}/\ptg< 0.07$ requirement in the main frame and their 
ratio in the lower frame. For comparison we show the same ratio at LO+PS 
and at NLO. 
The effects of the additional isolation are vanishingly small at LO+PS, 
which can be considered a lower bound for the impact that NG logarithmic terms stemming from photon isolation have on the all-order prediction of \ptllg{}.
The ratios at
NLO+PS and at NLO are very similar to each other, with the matching to PS slightly reducing the effects due the additional 
isolation requirement. Their difference can be regarded as an estimate of the size of the NG logarithmic 
corrections beyond fixed order induced by the $p_T^{\rm cone0.2}/\ptg< 0.07$ requirement. Since the difference is very small at low \ptllg{} and at most $\sim 2$\% in the matching region, we neglect such effect from now on.
We note that it is less straightforward to estimate the NG logarithmic contributions 
for the Frixione smooth-cone isolation, which for IR safety cannot be 
removed. However, we have verified that by varying the smooth-cone radius down to $\delta_0=0.01$ the analogous difference is only moderately affected and 
remains negligible at and below the peak of the spectrum.

We conclude our analysis by comparing our NNLO+N$^3$LL predictions
to $13$\,TeV ATLAS data \cite{Aad:2019gpq} in \fig{fig:atlas}. The analysis of \citere{Aad:2019gpq} is the first
diboson measurement that includes the full Run~II data set. 
The agreement is truly remarkable, especially 
with the precision of both theoretical prediction and data being at the few-percent level.
The shape of the distribution is very well
described by the predicted spectrum, and none of the data points is more than one standard deviation away from the theoretical uncertainty band.
We observe that resummation and matching are crucial not only at small \ptllg{}, but also in the intermediate region 
$40 \lesssim\ptllg\lesssim 200$\,GeV, where the comparison to data is significantly improved with respect to the NNLO comparison carried out in \citere{Aad:2019gpq}.
 Furthermore, our results are a clear improvement over the comparison against NLO+PS predictions 
in \citere{Aad:2019gpq}. In conclusion, our resummed results not only constitute the most precise prediction of the spectrum to date, but they also provide the most accurate 
description of the $13$\,TeV ATLAS data.

To summarize, we have presented the first calculation of the transverse-momentum spectrum of $Z\gamma$ pairs at NNLO+N$^3$LL.
At high transverse momenta we exploit the most accurate fixed-order prediction known to date, while at 
small transverse momenta we perform transverse-momentum resummation at N$^3$LL accuracy for the first time.
Furthermore, our matching approach respects the unitarity of the spectrum, so that its integral yields 
exactly the total cross section at NNLO.
Our results show that higher-order corrections in both the fixed-order and the logarithmic series are mandatory 
to obtain a reliable description of the distribution. Comparing NLO+NLL to NNLO+N$^3$LL predictions we find  
corrections of more than $30$\% in the tail of the distribution both in our inclusive and our fiducial setup. Those are inherited directly from the
large NNLO corrections. Around the peak of the spectrum, we find corrections between $10$\% and $20$\% with a clear 
change in shape of the distribution. Moreover, at NNLO+N$^3$LL the peak moves by $1$--$2$\,GeV towards larger transverse 
momenta with respect to NLO+NLL. 
The inclusion of higher-order corrections substantially reduces scale uncertainties, especially in the region of small transverse momenta.
Furthermore, by means of an NLO+PS simulation we have estimated the impact of including NG logarithmic contributions 
beyond fixed order and found it to be minor with respect to the NNLO+N$^3$LL scale uncertainties.
Finally, we have compared our best prediction at NNLO+N$^3$LL to 
ATLAS data at $13$\,TeV for the transverse-momentum spectrum of the $Z\gamma$ pair, and found a remarkable 
agreement within uncertainties at the few-percent level.
We reckon that our results will play a crucial role in the rich physics programme that is based on precision studies of $Z\gamma$ 
production at the LHC.

\noindent {\bf Acknowledgements.}
We are indebted to Pier Monni, Massimiliano Grazzini, Stefan Kallweit and Emanuele Re for stimulating discussions and comments on the manuscript. 
MW would like to thank Giulia Zanderighi and Daniele Lombardi for useful discussions.
The work of LR is supported by the ERC Starting Grant 714788 REINVENT.
\vspace{0.1cm}
\setlength{\bibsep}{3.5pt}
\renewcommand{\em}{}
\bibliographystyle{apsrev4-1}
\bibliography{zgam_matrix+radish}

\begin{thebibliography}{78}
\bibitem [{mun()}]{munich}%
  \BibitemOpen
  \href@noop {} {}\bibinfo {note} {{\sc Munich} is the abbreviation of
  ``MUlti-chaNnel Integrator at Swiss (CH) precision''---an automated parton
  level NLO generator by S.~Kallweit. In preparation.}\BibitemShut {Stop}%
\end{thebibliography}
\end{document}